\documentclass[12pt]{article}
\usepackage{latexsym,amsfonts,amssymb,amsmath}
\usepackage[english,russian]{babel}
\usepackage[utf8]{inputenc}
\usepackage{amssymb}
\usepackage{amsmath}
\usepackage[dvips]{graphicx}
\usepackage{epsfig}

\newcommand{\bab}{\end{gather}}
\newcommand{\ri}{{\mathrm i}}
\newcommand{\p}{\partial}
\newcommand{\bea}{\begin{array}}
\newcommand{\eea}{\end{array}}
\newcommand{\beg}{\begin{gather}}
\catcode`@=11 \long
\def\@caption#1[#2]#3{\par\addcontentsline{\csname
ext@#1\endcsname}{#1} {\protect\numberline{\csname
the#1\endcsname}{\ignorespaces #2}} \begingroup \small
\@parboxrestore \@makecaption{\csname fnum@#1\endcsname}
{\ignorespaces #3}\par \endgroup} \catcode`@=12

\newcommand{\la}{\label}

\catcode`@=11 \long
\def\@caption#1[#2]#3{\par\addcontentsline{\csname
ext@#1\endcsname}{#1} {\protect\numberline{\csname
the#1\endcsname}{\ignorespaces #2}} \begingroup \small
\@parboxrestore \@makecaption{\csname fnum@#1\endcsname}
{\ignorespaces #3}\par \endgroup} \catcode`@=12

\begin{document}
\allowdisplaybreaks

 \begin{titlepage} \vskip 2cm

\begin{center} {\Large\bf Superintegrable and scale invariant quantum mechanical systems with position dependent mass}

\vskip 3cm {\bf A. G. Nikitin}\footnote{E-mail: {\tt nikitin@imath.kiev.ua}}\vskip .5cm \rm \vskip 5pt {\sl Institute of Mathematics of National Academy of Sciences of Ukraine,\\3 Tereshchenkivska Street, Kyiv 4, Ukraine,
 01004\\}

\end{center}

\begin{center}\bf \large Abstract\end{center}

{Schr\"odinger equations with position dependent mass which are scale invariant and admit second order integrals of motion are classified.}

\vspace{10mm}

\end{titlepage}
\section{Introduction}

It is well known that fundamental equations of mathematical physics admit rather exten-ded symmetries. At the first place we can mention symmetries with respect to continuous groups (Lie symmetries). In addition, there exists an extended class of generalised symmet-ries (higher symmetries, supersymmetries, hidden symmetries)  connected with the menti-oned equations, see, e.g., book
 \cite{FN1} and  paper \cite{FN11}.

An important class of equations of quantum mechanics (QM) is formed by the superin-tegrable Schr\"odinger equations which admit more integrals of motion than the number of degrees of freedom of the associated QM systems. As a rule such equations also admit a wide Lie symmetry and different kinds of generalised symmetries. A known example is the Schr\"odinger equations for the Hydrogen atom which, in addition to its invariance with respect to the rotation group, is supersymmetric and admits the hidden (Fock) symmetry with respect to group O(4).

Searching for Lie symmetries of partial differential equations is a rather popular business which generates a big number of publications. Rather surprisingly, the correct group classification of Schr\"odinger equations for a particle interacting with an external field appears only recently \cite{N22, N32, N42, N52}, in spite of that the search for the related symmetries started long time ago, see papers \cite{Nied, And, Boy}.

The systematic search for superintegrable QM systems was started in papers of Yakov Smorodinsky and his collaborators  \cite{wint1}, \cite{BM}, who completely described all inequivalent 2d Schr\"odinger equations which admit second order integrals of motion. The modern trend is the studying  of the related integrals of motion of the third and even arbitrary orders \cite{Mar}. See also  \cite{AGN1} where the determining equations for such symmetries were deduced.

The present paper is devoted to the classification of superintegrable Schr\"odinger equa-tions with position dependent mass (PDM). Such equations have  very important applica-tions in the modern theoretical physics, being the mathematical models of semiconductors \cite{Roz}, quantum liquids  \cite{7},  quantum dots \cite{3}, and many, many others quantum objects.

Higher symmetries of the 2d PDM Schr\"odinger equations are well known  (see, e.g.,  \cite{Miller1} and the references given therein). However, there are only some particular results concerning the classification of such equations with three independent variables  \cite{Miller2, Bala2, Rag1, N62}. In the following we present the complete classification of the special class of such equations, namely, of equations which are scale invariant.

For the group classification of the stationary and time dependent PDM Schr\"odinger equations and the related reaction-diffusion systems see papers  \cite{NZ, AGN} and \cite{AGN0, NNN} corres-pondingly. In accordance with these papers such equations cannot be Galilei invariant in contrast  with the standard Schr\"odinger equation and its generalisations for the case of particles with arbitrary spin \cite{NNNN}.

The classification of the higher symmetries of the 3d PDM Schr\"odinger equations is a very complicated problem which is solved only for some special classes of such equations. An important and rather extended class of them includes equations which admit at least one parametric symmetry group. All inequivalent equations from this class which admit first order integrals of motion is fixed in \cite{NZ}.

In the present paper we classify the scale invariant 3d  PDM Schr\"odinger equations which  admit at least one second order integral of motion and show that they include many interesting and consistent equations. The found equations  form the important part of a more general class of superintegrable equations which admit at least a minimal Lie symmetry.

\section{PDM Schr\"odinger equations}

We will study the stationary PDM Schr\"odinger equations of the following generic form:
\begin{gather}\la{se}
   \hat H \psi=E \psi,
\end{gather}
where
\begin{gather}\la{H}\hat H=p_af({\bf x})p_a+ V({\bf x}).\end{gather}
In equation (\ref{se}) ${\bf x}=(x^1,x^2,x^3),$ $p_a=-i\p_a$, а $V({\bf x})$ та  $f({\bf x})=\frac1{2m({\bf x})}$ are arbitrary functions of  ${\bf x}$ which are associated with the potential and inverse mass correspondingly, and summation is imposed over the repeated indices.

In paper  \cite{NZ} all inequivalent equations (\ref{se}) admitting at least one first order integral of motion (i.e., differential operator of first order commuting with Hamiltonian $H$) are found. In  \cite{N62} the complete classification of the special class of superintegrable equations (\ref{se}) was carried out, namely, the equations invariant with respect to the rotation group.

In the present paper we classify superintegrable equations (\ref{se}) which are scale invariant. The corresponding arbitrary elements  $V$ and  $f$ take the form  \cite{NZ}:
\begin{gather}\la{f_V1}f= r^2F(\varphi,\theta),\quad V=V(\varphi),\end{gather}
where $F(.)$ and $V(.)$ are arbitrary functions, $\varphi$ і  $\theta$ are the Euler angles.

Hamiltonians (\ref{H}) with special arbitrary elements (\ref{f_V1}) commute with the generator of the scaling transformations which has the following form:
\begin{gather*} D=x_ap_a-\frac{3\ri}{2}.\end{gather*}
Our task is to find such of them which commute with at least one first order differential operator.

\section{Equivalence group}

Changes of dependent and independent variables are called the equivalence transformati-ons provided they keep the generic form of the differential equation (in our case of equation  (\ref{se})) up to the changes of the arbitrary elements  (in our case functions $f$ and $V$). The set of the equivalence transformations has the gruppoid structure \cite{Rom}, and can include equivalence groups and some discrete elements.

As it is shown in \cite{NZ}, the maximal continuous equivalence group of equation  (\ref{se}) is C(3), i.e., the group of conformal transformations of 3d Euclidean space. The generators of this group are the following first order differential operators:
\begin{gather}\label{QQ}\begin{split}&
 P^{a}=p^{a}=-i\frac{\partial}{\partial x_{a}},\quad L^{a}=\varepsilon^{abc}x^bp^c, \\&
D=x_n p^n-\frac{3\ri}2,\quad K^{a}=r^2 p^a -2x^aD,\end{split}
\end{gather}
where $r^2=x_1^2+x_2^2+x_3^2$  and $p_a=-i\frac{\p}{\p x_a}.$
The corresponding group transformations (whose explicit form can be found in \cite{NZ}) keep the generic form of equations  (\ref{se}), (\ref{H}) up to exact form of functions $f$ and $V$. The important particular form of these transformations is the inversion:
 \begin{gather}\la{IT} x_a\to
\tilde x_a=\frac{x_a}{x^2},\quad \psi({\bf x})\to \tilde x^3\psi(\tilde{\bf x})\end{gather}
which acts on operators  (\ref{QQ}) in the following manner:
\beg P_a\to K_a,\quad K_a\to P_a, \quad L_a\to L_a, \quad D\to D.\la{inv}\end{gather}

For the class of equations considered in the present paper the equivalence group is reduced to the direct product of the rotations group whose generators are componets of the orbital momenta $L_a$ and dilatation transformations generated by $D$, since $L_a$ and $D$ commute with $H$ while the remaining operators  (\ref{QQ}) do not have this property. Notice that the discrete equivalence transformation (\ref{IT}) is kept also.

In the following we will use the rotations and the inverse transformation  (\ref{IT}) for optimisation of the requested  calculation and selection of non-equivalent versions of the studied equations.

\section{Determining equations}
The searched second order integrals of motion can be represented in the following form:
\begin{equation}\label{Q}
    Q=\mu^{ab}\p_a\p_b+\xi^a\p_a+\eta
\end{equation}
where $\mu^{ab}=\mu^{ba}$, $\xi^a$ and $\eta$ are unknown functions of  $\bf
x$.

By definition operators    $Q$ have to commute  $\hat H$:
\begin{equation}\label{HQ}[ \hat H,Q]\equiv  \hat H Q-Q \hat H=0.\end{equation}
Condition  (\ref{HQ}) represents the operator equation which can be satisfied when the operators presented here act on an arbitrary twice differentiable function. Evaluating the commuta-tor and equating the coefficients for the same differential operators $\frac{\p}{\p x_a}, \frac{\p^2}{\p x_a\p x_b}$ and $ \frac{\p^3}{ \p x_a\p x_b\p x_c}$ we come to the following system of determining equations:
\begin{gather}\la{mmmu0}5\left(\mu^{ab}_c+\mu^{ac}_b+ \mu^{bc}_a\right)=
\delta^{ab}\left(\mu^{nn}_c+2\mu^{cn}_n\right)+
\delta^{bc}\left(\mu^{nn}_a+2\mu^{an}_n\right)+\delta^{ac}
\left(\mu^{nn}_b+2\mu^{bn}_n\right),\\
\la{mmmu1}
 \left(\mu^{nn}_a+2\mu^{na}_n\right)f-
5\mu^{an}f_n=0,\\
2f\eta_a+\xi^a_{n}f_n-\xi^nf_{an}+f\xi^a_{nn}+2\mu^{an}V_n-
\mu^{mn}f_{mna}=0,\la{mmmu4}
\\
\left(\mu^{ab}_{nn}+\xi^a_b+\xi^b_a\right)f+
\mu^{ab}_nf_n-\mu^{na}f_{nb}-\mu^{nb}f_{na} -\delta^{ab}
\left(\mu^{mn}f_{mn}+\xi^nf_n\right)=0,\la{mmmu2}\\
 f\left(\mu^{mm}_{nn}+2\xi^n_n\right) +\left(\mu^{nn}_m-3\xi^m\right)f_m-5\mu^{mn}f_{mn}=0,
\la{mmmu3}\\
 (f\eta_n)_n+\xi^nV_n+\mu^{mn}V_{mn}=0\la{mmmu5}
\end{gather}
  where $f_n=\frac{\p f}{\p x_n}, \ \xi^a_n=\frac{\p \xi^a}{\p x_n}$, etc.

To find all operators (\ref{H}) which admit integrals of motion  (\ref{Q}) it is necessary to find all inequivalent solutions of the very complicated system of equations (\ref{mmmu0})--(\ref{mmmu5}) for twelve unknowns $\mu^{ab}, \xi^a, \eta, f$ and  $V$.
Fortunately, equations  (\ref{mmmu2}), (\ref{mmmu3}) and (\ref{mmmu5}) can be omitted since they are nothing but the differential consequences of the remaining ones. In addition, variables , $\xi^a$ can be excluded. Indeed, diffirentiating  (\ref{mmmu1}) with respect to  $x_a$ and making the summation over the repeating index $a$,  we come to equation
 (\ref{mmmu3}), while the same procedure with   (\ref{mmmu4}) lead to equation (\ref{mmmu5}). On the other hand, equation  (\ref{mmmu3}) can be deduced from
(\ref{mmmu0}) and (\ref{mmmu1}), provided the following relations are satisfied:
\begin{gather}\begin{split}\label{de1}&\hat \xi^b_{a}+\hat \xi^a_{b}=\frac23\delta_{ab}\hat\xi^n_n,\\&
3\hat\xi^nf_n=2f\hat\xi^n_n\end{split}\end{gather} where $\hat \xi^a=\xi^a-\mu^{an}_n$ .

Relations  (\ref{de1}) are nothing but the determining equations for the coefficients of the first order symmetry operators of equations (\ref{se}), found in paper \cite{NZ}. Since such operators and the corresponding equations (\ref{se}) are known \cite{NZ}, we suppose that functions  $\hat \xi^a$ are trivial, and so  $\xi^a=\mu^{an}_n$. As a result equations   (\ref{mmmu4})  are reduced to the following form:
\begin{gather}\la{mmmu6}2(f\eta^a+\mu^{ab}V_b)+(\mu^{am}_{nm}f-\mu^{nm}f_{am})_n=0, \end{gather} and the corresponding integral of motion   (\ref{Q}) can be rewritten as:
  \begin{gather}\label{Q1}
    Q=\p_b\mu^{ab}\p_a+\eta.
\end{gather}

The corresponding functions   $\mu^{ab}$ have to satisfy the autonomous system of equations (\ref{mmmu0}), which specify the conformal Killing tensor.  The latter one is a linear combination of the following tensors
(see, e.g., \cite{Kil}):
\begin{gather}\la{mmu1}\begin{split}& \mu^{ab}_1=\lambda_1^{ab}+\delta^{ab}\frac{\lambda_1^{cd}x^cx^d}{x^2}\varphi_1(x),
\\&
\mu^{ab}_2=\lambda_2^a x^b+\lambda_2^b x^a+\delta^{ab}\lambda_3^c x^c\varphi_2(x),
\\&
\mu^{ab}_3=(\varepsilon^{acd}\lambda_3^{cb}+ \varepsilon^{bcd}
\lambda_3^{ca})x^d, \\&
\mu^{ab}_4=(x^a\varepsilon^{bcd}+x^b\varepsilon^{acd}) x^c\lambda_4^d,
\\&\mu^{ab}_5=\delta^{ab}x^2\varphi_5(x)+
k (x^ax^b-\delta^{ab}x^2),\\&
\mu^{ab}_6=\lambda_5^{ab}x^2-(x^2\lambda_5^{bc}+x^b\lambda_5^{ac})x^c+
\delta^{ab}\lambda_6^{cd}x^cx^d\varphi_6(x),\end{split}\\\begin{split}
& \mu^{ab}_7=(x^a\lambda_7^b+x^b\lambda^a)x^2-4x^ax^b\lambda_7^c x^c+
\delta^{ab}
\lambda_8^c x^cx^2\varphi_7(x),
\\&\mu^{ab}_8= 2(x^a\varepsilon^{bcd} +x^b\varepsilon^{acd})
\lambda_8^{dn}x^cx^n- (\varepsilon^{ack}\lambda_8^{bk}+
\varepsilon^{bck}\lambda_8^{ak})x^cx^2,\\&
\mu^{ab}_9=\lambda_9^{ab}x^4-2(x^a\lambda_9^{bc}+x^b\lambda_9^{ac})x^cx^2+
(4x^ax^b+k\delta^{ab}x^2)\lambda_{9}^{cd}x^cx^d\\&+\delta^{ab}\lambda_{10}^{cd}x^cx^dx^2
\varphi_9(x)\end{split}\la{mmu2}
\end{gather}
where $\lambda_n^{ab}=\lambda_n^{ba}$ and  $\lambda_n^a $  are arbitrary parameters, and  $\varphi_1,...,\varphi_9$ are arbitrary functions of
$x=r=\sqrt{x_1^2+x_2^2+x_3^2}$.

Thus our classification problem is reduced to finding inequivalent solutions of equations (\ref{mmmu1})
and (\ref{mmmu6}) for unknowns  $f$  and $V$, where   $\mu^{ab}$ is a linear combination of functions   (\ref{mmu1}) and (\ref{mmu2}). The main difficulty is the extended number of arbitrary parameters involved into the mentioned equations which has to be reduced using the equivalence transformations.

\section{Decoupling of the determining equations}

 Taking into account the scale invariance of equations (\ref{se}) with arbitrary elements (\ref{f_V1}) we can conclude that the Killing tensors $\mu^{ab}$ involved into the determining equations should be homogeneous functions of $\bf x$. In other words, equations ,   (\ref{mmmu1})
and (\ref{mmmu6}) are decoupled to five autonomous subsystems which correspond to the Killing tensors including homogeneous polynomials of fixed order $n$ with $n=0, 1, 2, 3, 4$ , and the corresponding functions $\varphi_1, \varphi_2, ...,\varphi_9$ should be constants.

Moreover since any equation  (\ref{se}), (\ref{H})   which arbitrary element  (\ref{f_V1}) is invariant w.r.t. the inverse transformation (\ref{IT}), we can restrict ourselves to the Killing tensors being polynomials of $x_a$ of order
 $n \leqslant2$, which are given by equations   (\ref{mmu1}), since the symmetries with   $n$=3 and $n$=4 are equivalent to the symmetries with $n=1$ and $n=0$ correspondingly. It means that it is sufficient to solve the determining equations (\ref{mmmu1})
and  (\ref{mmmu6}) with the following  $\mu^{ab}$:
\begin{gather}\la{m0}\mu^{ab}=\lambda^{ab}+
\kappa\delta^{ab}\frac{\tilde\lambda^{cd}x^cx^d}{x^2},\\\la{mm1}\begin{split}&
\mu^{ab}=\lambda^a x^b+\lambda^b x^a-2\delta^{ab}\tilde\lambda^cx^c \\&+\mu^ax^b+\mu^bx^a +(\varepsilon^{acd}\lambda^{cb}+ \varepsilon^{bcd}
\lambda^{ca})x^d,\end{split}\\\la{mm2}\begin{split}&
\mu^{ab}=\kappa x^ax^b+(x^a\varepsilon^{bcd}+x^b\varepsilon^{acd})\lambda^dx^c+
\delta^{ab}\tilde\lambda^{cd}x^cx^d\\&+
\lambda^{ab}x^2-(x^a\lambda^{bc}+x^b\lambda^{ac})x_c.\end{split}\end{gather}
Notice than any second order symmetry corresponding to $n=0$ and $n=1$ is accompanied by the addition symmetry generated by the changes of variables  (\ref{inv}).
\subsection{Symmetries independent on $x_a$}
Let us start with the symmetries which correspond to Killing tensors $\mu^{ab}$ fixed in (\ref{m0}). Substituting
(\ref{m0}) into  (\ref{mmmu3}), we obtain the following equation:
\begin{gather}\la{deq1} \lambda^{ab}f_b+\kappa \frac{\tilde\lambda^{mn}x_mx_n}{r^2}f_a=0.\end{gather}

Since  $\mu^{ab}=\mu^{ba}$, up to rotation transformations there are three inequivalent versions of nontrivial parameters  $\lambda^{ab}$:
\begin{gather}\la{deq111}\lambda^{11}=k_1, \lambda^{22}=k_2, \lambda^{33}=k_3,\\\la{deq121}\lambda^{11}=k_1, \lambda^{22}=k_2,\\\la{deq131}\lambda^{33}=k_3.\end{gather}
Substituting  (\ref{deq111})-(\ref{deq131}) into (\ref{deq1}) and
equating the coefficients for the same independent variables   $x_m$, we come to the conclusion that coefficient $\kappa$ should be trivial, and
\begin{gather}\la{deq14}f=0 \ \    \text{ for version }   (\ref{deq111}), \\\la{deq15}f=x_3^2 \ \       \text{ for version  }   (\ref{deq121}),
 \\\la{deq16}f=(x_1^2+x_2^2)F(\varphi) \ \       \text{  for version  }   (\ref{deq131})\end{gather}
 where $F(\varphi)$ is an arbitrary function of the Euler angle $\varphi=\arctan\left(\frac{x_2}{x_1}\right).$

Substituting    (\ref{deq11})-(\ref{deq16}) into  (\ref{mmmu6}) we obtain the following solutions for the latter equation:
\begin{gather}\la{sol1}f=x_3^2, \quad V=\frac{x_3^2}{(x_1^2+x_2^2)}F(\varphi), \quad  \eta=-\frac1{x_1^2+x_2^2}F(\varphi),\quad  \lambda^{11}=\lambda^{22}=1,\\\la{sol2}f=x_3^2,\quad V=Const,\quad  \eta=0, \quad \lambda^{11}=k_1,\ \lambda^{22}=k_2\neq k_1, \\\la{sol3}f=(x_1^2+x_2^2)F(\varphi),\quad V=\frac{f}{x_3^2},\quad \eta=-\frac1{x_3^2},\quad \lambda^{33}=1.\end{gather}

Thus up to the equivalence there are three versions of the scale invariant PDM Schr\"odin-ger equations which admit second order integrals of motion independent on $\bf x$. They include arbitrary elements $f$ and  $V$  fixed in formulae (\ref{sol1}) and (\ref{sol2}).

\subsection{Symmetries linear in independent variables}

Consider now the symmetries generated by the Killing tensors (\ref{mm2}) which are linear in $x_a$. The related determining equations  (\ref{mmmu3}) are reduced to the following form:
\begin{gather} \la{deq4}2\lambda^af=\mu^{ab}f_b.\end{gather}

To simplify calculations we will use the identity   $2f=x_1f_1+x_2f_2+x_3f_3$ which makes it possible to reduce  (\ref{deq4}) to the following    {\it homogeneous} system of linear algebraic equations for deraviatives   $f_a$:
\begin{gather}\la{deq5}M^{ab}f_b=0,\end{gather}
where
\begin{gather}\la{deq6} M^{ab}=\mu^{ab}-\lambda^ax^b-\mu^ax_b.\end{gather}

Notice that operator (\ref{Q}) with   $\mu^{ab}$ given in  (\ref{sol2}) is a bilinear combination of generators of conformal group , $C(3)$, i.e.,
\begin{gather}\la{lc}\begin{split}&Q=a_+P_2L_3+a_-L_2P_3+b_+P_1L_3+b_-P_3L_1+c_+P_1L_2\\
&+c_-P_2L_1+\lambda^ap_a D+\tilde d_1P_1L_1+\tilde d_2P_2L_2+\tilde d_3L_3P_3,\end{split}\end{gather}
where $a_\pm=\lambda^{23}\pm\lambda^1, b_\pm=\lambda^{31}\pm\lambda^2, c_\pm=\lambda^{12}\pm \lambda^3, \tilde d_a=\lambda^{bb}-\lambda^{cc}$, (a, b, c) is the cycle (1, 2, 3).

Using rotation transformations coefficients  $a_1, b_1$ and $c_2$ can be reduced to zero. As a result the components of tensor  $M^{ab}$ (\ref{deq6}) take the following form:

 \begin{gather}\la{deq7}\begin{split}&M^{11}=-2cx_3+\lambda_1x_1,\ M^{12}=\lambda_2x_1+d_3x_3,
  M^{21}=\lambda_1x_2+ d_3x_3,\\& M^{22}=\lambda_2x_2, \ M^{23}= d_1x_1+\lambda_3x_2+bx_3,
  \\& M^{13}=ax_3+d_2x_2+\lambda_3x_1, \  M^{31}=cx_1+ d_2x_2+(a+\lambda_1)x_3,\\& M^{32}=(b+\lambda_2)x_3+ d_1x_1, \ M^{33}=\lambda_3x_3-2ax_1-2bx_2
 \end{split}\end{gather}
 where we redenote $a_-=a,\ b_-=b,\ c_+=c,\ d_1=\tilde d_2-\tilde d_3, \ d_2=\tilde d_3-\tilde d_1,\ d_3=\tilde d_1-\tilde d_2$.

Equation (\ref{deq5}) admits nontrivial solution iff the determinant of the matrix whose entries $M^{ab}$ are given by equations
(\ref{deq7}) is equal to zero It is the case provided one of the following conditions is satisfied:
\begin{gather}\la{det4}a=b=c=d_a=0,\\\la{det6}d_1=-d_2, c=0, \ \lambda_1=\lambda_2=0,\\\la{det1}ad_2=-bc, \ b\lambda_1 =-2\lambda_3,\\\la{det2}b\lambda_1=a\lambda_2, \ c=0, \\\la{det3}a=b=d_a=\lambda_2=0,\\\la{det5}\lambda_1d_1=-c\lambda_2, \ d_1\lambda_3=-a\lambda_2,\ b=0.\end{gather}

Substituting consequently the versions presented in  (\ref{det4})-- (\ref{det5}) into equations  (\ref{deq5}) and (\ref{mmmu6}) with matrices   (\ref{deq7}) we find the corresponding functions $f$, $V$ and $\eta$.

Let us start with version (\ref{det4}). In this case the symmetry operator includes a linear combination of generators $P_1D, P_2D$ and $P_3D$ with coefficients  $\lambda_1, \lambda_2$ and $\lambda_3$ correspon-dingly. Rotating the coordinate system, two of these coefficients (say, $\lambda_1$ and $\lambda_2$) can be nullified which reduce the tensor components given in   (\ref{deq7}) to the following form:
\begin{gather*}M^{3a}=x_a,\ a=1, 2, 3,\end{gather*}
while the remaining components are reduced to zero. The related equations   (\ref{deq5}) and (\ref{mmmu6}) take the following form:
\begin{gather}\la{m1} x_af_3=0, \\\la{m2}f\eta_a+x_aV_3=0.\end{gather}

The generic solution of equation (\ref{m1})  for function $f$  given in  (\ref{f_V1})  is:
\begin{gather}\la{m3}f=\tilde r^2F(\varphi),\end{gather}
where $F(\varphi)$  is an arbitrary function of the Euler angle. Integrating the related equation (\ref{m2}) for  $V$ given in (\ref{f_V1}), we find the generic form of functions  $\eta$ and $V$:
\begin{gather*}V=G(\varphi)+c_1\frac{x_3} {\tilde r}F(\varphi)+c_2\frac{x_3}rF(\varphi)\\
\eta=\frac{c_1}{\tilde r}+\frac{c_2}{r}\end{gather*}
where $c_1$ and $c_2$  are integration constants, and  $G(\varphi)$ is an additional arbitrary function.

The found solutions are represented in the firs item of Table  1, where the additional integral of motion obtained by the inversion transformation is fixed also.

Considering version (\ref{det6}) we come to the following nontrivial components of tensor
 $M^{ab}$ (\ref{deq7}):
 \begin{gather*} M^{13}=ax_3+dx_2+\lambda_3x_1, \ M^{23}=\lambda_3x_2-dx_1-bx_3,\\M^{31}=ax_3+dx_2,\ M^{32}=-dx_1-bx_3,\ M^{33}= \lambda_3x_3+2bx_2-2ax_1.\end{gather*}
 Up to rotation transformations we can set $b=0$,
and the related equations  (\ref{deq5}) are reduced to the following ones:
\begin{gather}\la{m5} f_3=0,\quad af_1=0, \quad d(x_1f_2-x_2f_1)=0.\end{gather}

In accordance with (\ref{m5}) one out of two coefficients  $a$ and $d$  should be trivial provided the mass is not a constant. The corresponding solutions of equation  (\ref{m5}) look as:
\begin{gather}\la{m6} f=x_2^2,\  \text{}   d=0, \ a\neq 0,\\ \la{m7}
                         f= \tilde r^2,\ \text{} a=0,\ d\neq 0.\end{gather}

In the cases (\ref{m6}) and (\ref{m7}) equations (\ref{mmmu6}) take the following forms:
\begin{gather}\la{m81}\begin{split}&x_2^2\eta_1+(\lambda_3x_1+ax_3)V_3=0,\\&x_2^2\eta_2
+\lambda_3x_2V_3=0,\\&x_2^2\eta_3+ax_3V_1+(\lambda_3x_3-2ax_1)V_3=0\end{split}\end{gather}
and
\begin{gather}\la{m8}\begin{split}&\tilde r^2\eta_1=-(dx_2+\lambda_3x_1)V_3,\\&\tilde r^2\eta_2=(dx_1-\lambda_3x_2)V_3,\\&\tilde r^2\eta_3=d(x_1V_2-x_2V_1)-\lambda_3x_3V_3.\end{split}\end{gather} correspondingly.

If all arbitrary parameters $a, \lambda_3$ and $d$ are nontrivial, equations (\ref{m81}) and (\ref{m8}) have only constant solutions for $V$ and $\eta$. Equations (\ref{m81})  for $a=0$ and equations (\ref{m8}) for $d=0$ are reduced to the particular cases of  equations (\ref{m2}) and (\ref{m3}).
Their solutions are represented in Items 2 of Table 1 where $F(\varphi)=0$ for  the system (\ref{m8}) and $F(\varphi)=\sin(\varphi)$ for  the system (\ref{m8}).

Obtained in analogous manner other versions of  Hamiltonians (\ref{H}) which admit second order integrals of motion linear in the independent variables are collected in Tables 1 and 2 where the systems which admit more than one integral of motion are indicated also.
\section{Symmetries bilinear in independent variables}

Le last task is to find PDM Schr\"odinger equations admitting integrals of motion which are generated by tensors (\ref{mm2}). These integrals of motion can be represented as:
\begin{gather}\la{Lc}Q=\nu^{ab}(\{K_a,P_b\} +\{P_b,K_a\})+\tilde \lambda^{ab}Q^{ab},\end{gather}
where $\nu^{ab}=\lambda^{ab}+\varepsilon_{abc}\lambda_c,$ $\tilde \lambda^{ab}$ are arbitrary koefficients,  $Q^{ab}=P_cx_ax_bP_c$, and the symbols $\{.,.\}$ denote anticommutators.

Operators $Q^{ab}$ also can be expressed via operators  (\ref{QQ}), since  the following identities are true:
\begin{gather}\la{iden}\begin{split}&\{L_a,L_b\}+\{P_a,K_b\}=2Q^{ab},\quad  a\neq b,\\&
\{P_1,K_1\}+\{ P_2,K_2\}+L_3^2=2Q^{33}.\end{split}\end{gather}

 The corresponding determining equations  (\ref{mmmu3}) take the following form:
\begin{gather} \la{deq41}2f(\lambda^{ab}-\tilde\lambda^{ab})x_b)=\mu^{ab}f_b.\end{gather}

In analogy with  (\ref{deq4})-(\ref{deq6})  it is convenient to rewrite equations  (\ref{deq41}) in the form  (\ref{deq5}), where
\begin{gather} \la{deq51}M^{ab}=\mu^{ab} -\lambda^{ac}x_cx_b.\end{gather}

Like in the case of equation  (\ref{deq7}), using the rotation transformations many of arbitary coefficients in (\ref{deq51}) can be nullified and the coefficient matrix can be reduced to one of the canonical forms.The related matrices  (\ref{deq51}) are reduced to linear combinations of the following matrices  $N^{ab}$ and $\tilde N^{ab}$:
\begin{gather}\la{deq11}M=\nu^{ab}N^{ab}+\tilde\nu^{ab}\tilde N^{ab},\end{gather}
where $\tilde N^{ab}=x_ax_b I$, $I$ is the unit matrix, and the nontrivial entries    $N^{ab}_{cd} $  of matrices $N^{ab} $ are given in the following formulae (refer to  (\ref{mm2}):
\begin{gather}\begin{split}&N_{11}^{12}=-x_1x_2, \ N_{11}^{13}=-x_1x_3, \ N_{11}^{22}=N_{11}^{33}=x_1^2,\\&N_{22}^{11}=N_{22}^{33}=x_2^2, \ N_{22}^{21}=-x_2x_1,\ N_{22}^{23}=-x_2x_3,\\&N_{33}^{31}=-x_3x_1,\ N_{33}^{32}=-x_3x_2,\  N_{33}^{33}=\tilde r^2,\\&N^{12}_{11}=-2x_1x_2,\ N^{12}_{12}=r^2,\  N^{12}_{21}=r^2-2x_2^2, \ N^{12}_{31}=-2x_2x_3,\\&N^{21}_{12}=r^2-2x_1^2,\ N^{21}_{21}=r^2,\ N^{21}_{22}=-2x_1x_2,\ N^{21}_{32}=-2x_1x_2.\end{split}\la{ono}\end{gather}

Matrices  (\ref{ono}) are degenerated and so the corresponding equations  (\ref{deq5}) and (\ref{mmmu6}) have nontrivial solutions. Inequivalent versions of these matrices are presented in the following formulae:
 \begin{gather}\la{ono1}N=\nu_1N^{11}+\nu_2N^{22}+\nu_3N^{33},\\
 \la{ono2}N=\nu_4(N^{11}+N^{22})+\nu_5N^{33}+\nu_6N^{12},\\
 \la{ono3}N=\nu_{7}(N^{11}+N^{22}+N^{33})+\nu_{8}N^{12}+\nu_{9}N^{23},\\
 \la{ono4}N=\nu_{10}(N^{11}+N^{22})+\nu_{11}N^{33}+\nu_{12}(N^{12}-N^{21}),\\
 \la{ono5}N=\nu_{13}(N^{11}+N^{22}+N^{33})+\nu_{14}(N^{12}-N^{21})+\nu_{15}(N^{23}-N^{32})
  \end{gather}
where $\nu_1, \nu_2, ..., \nu_{15}$  are arbitrary linear coefficients.

Thus the classification of the integrals of motion bilinear in  $x_a$  is reduced to search of generic solutions of equations (\ref{deq5}) and (\ref{mmmu6}), where  $M$ are matrices given by formulae (\ref{deq11}) and
 (\ref{ono1})-(\ref{ono5}).
  We will not present the details of the requested cumbersome calculations but restrict ourselves to one particular example of them.

 Let matrix  $M$ is reduced to (\ref{ono4}) with nontrivial  $\nu_{10}$  and $\nu_{12}$. Then equations    (\ref{deq5}) have nontrivial solutions  $f$ if $\nu_{11}=0$, and equations  (\ref{mmmu6}) have nontrivial solutions for $V$ if  $\nu_{10}=\nu_{12}$ or $\nu_{10}\nu_{12}=0.$

 If  $\nu_{10}=\nu_{11}=0$ then equations (\ref{deq5}) and  (\ref{mmmu6}) take the following form:
 \begin{gather*}x_1f_2-x_2f_1=0\end{gather*}
and
\begin{gather*}f\eta_a=-2x_a(x_1V_2-x_2V_1), \ a=1, 2, 3 \end{gather*}
correspondingly. Their generic solutions look as:
\begin{gather*}f=r^2F(\theta),\ V=cF(\theta)\varphi+G(\theta),\ \eta=-c\ln(r^2). \end{gather*}

If $\nu_{10}\neq 0, \nu_{12}=\nu_{11}=0$ we have the following versions of equations  (\ref{deq5}) and  (\ref{mmmu6}):
\begin{gather*}\begin{split}& f_1=0, f_2=0,\\&
 f\eta_1+(x_2^2+x_3^2)V_1-x_1x_2)V_2=0,\\& f\eta_2-x_1x_2V_1+(x_1^2+x_3^2)V_2=0,\\& f\eta_3-x_1x_3V_1-x_2x_3V_2=0,\end{split}\end{gather*} and the related solutions are:
 \begin{gather*}f=x_3^2,\ V=\frac{x_3^2}{\tilde r^2}F(\varphi)+G(\theta),\ \eta=-\frac{r^2}{\tilde r^2}F(\varphi)-G(\theta).\end{gather*}

  \begin{center}Table 1. Inverse masses, potentials  and the related integrals of motion defined up to arbitrary functions.\end{center}
\begin{tabular}{c c c c}

\hline

\vspace{1.5mm}

 №&$f$&$V$&\text{Integrals of motion}\\
\hline\\
1\vspace{1.5mm}&$\tilde r^2F(\varphi)$&$F(\varphi)G(\theta)$&$\{P_3,K_3\}+4G(\theta)$\\
2\vspace{1.5mm}&$\tilde r^2F(\varphi)$& $c F(\varphi)\frac{x_3}r+G(\varphi)$&$\begin{array}{c}\{P_3,D\}-\frac{2c}{r}, \
\{K_3,D\}-{2c}{ r} \end{array}$\\
3\vspace{4mm}&$\tilde r^2F(\varphi)$& $c F(\varphi)\frac{x_3} { r}$&$\begin{array}{c}\{P_3,K_3\}+\frac{4c}{r}, \ \{P_3,D\}-\frac{2c}{r}, \\
\{K_3,D\}-{2c}{ r}\end{array}$\\
4\vspace{1.5mm}&$\tilde r^2F(\varphi)$& $F(\varphi)$&$\begin{array}{c}P_1^2+P_2^2+\frac{1}{ x_3^2}F(\varphi),\\K_1^2+K_2^2+\frac{r^4}{x_3^2}, \\ \{P_3,K_3\},\ \{P_3,D\}, \ \{K_3,D\}\end{array}$\\
5\vspace{1.5mm}&$\tilde r^2F(\varphi)$&$\frac{\tilde r^2}{x_3^2}F(\varphi)$&$\begin{array}{c}P_3^2+\frac1{x_3^2},\ K_3^2+\frac{r^4}{x_3^2},\\ \{P_3,K_3\}+\frac{4\tilde r^2}{x_3^2}\end{array}$\\
6\vspace{1.5mm}&$\tilde r^2F(\theta)$&$G(\varphi)F(\theta)+R(\theta)$&$\begin{array}{c}L_3^2+2G({\varphi})\end{array}$\\
7\vspace{1.5mm}&$ r^2F(\theta)$&$cF(\theta)\varphi+G(\theta)$&$\begin{array}{c}
\{L_3,D\}+2c\ln(r)\end{array}$\\
8\vspace{1.5mm}&$x_3^2$&$\frac{x_3^2}{\tilde r^2}F(\varphi)+G(\theta)$&$\begin{array}{c}\{P_1,K_1\}+\{P_2,K_2\}\\+\frac{4r^2}{\tilde r^2}F(\varphi)+4G(\theta),\\L_3^2+2F({\varphi})\end{array}$\\
9\vspace{1.5mm}&$x_3^2$&$c\frac{x_3^2\varphi}{ r^2}+G(\theta)$&$\begin{array}{c}
\{P_1,K_1\}+\{P_2,K_2\}+4\frac{r^2}{\tilde r^2}\varphi+4G(\theta),\\L_3^2+2{\varphi},\ \{L_3,D\}+c\ln(r)\end{array}$\\
10\vspace{1.5mm}&$x_2^2$&$F(\varphi)-\frac{x_1^2}{\tilde r^2}G(\theta)$&$\begin{array}{c}\{K_1,P_1\}-2J_2^2+\frac{4x_1^2}{\tilde r^2}G(\theta)-4F(\varphi),\\ \{K_3,P_3\}+4G(\theta)\end{array} $\\
11\vspace{1.5mm}&$x_3^2$&$\tilde V$&$\begin{array}{c}\{P_1,K_1\}+(b+1-a)\{P_2,K_2\}\\+2(a-1)J_3^2+4\tilde \eta
\end{array}$\\
\hline \hline
\end{tabular}

Finally, if  $\nu_{10}=\nu_{12}$, $\nu_{11}=0$, we have the following equations  (\ref{deq5}) and  (\ref{mmmu6}):
 \begin{gather*}f_1=0, \ f_2=0,\\f\eta_1+(x_2^2+x_3^2-2x_1x_2)V_1+(2x_1^2-x_1x_2)V_2=0,\\
 f\eta_2-(x_1x_2+2x_2^2)V_1+(x_1^2+2x_1x2+x_3^2)V_2=0,\\
 f\eta_3-(x_1x_3+2x_2x3)V_1+(2x_1x_3-x_2x_3)V_2=0,\end{gather*}
 and the related solutions are $f=x_3^2, \ V=F(\theta),\ \eta=-F(\theta)$.

Consider also the most complicated case when the symmetry operator (\ref{Lc}) is reduced to the following bilinear form of generators of group C(3):
\begin{gather}\la{bc}Q=\{K_1,P_1\}+\mu\{K_2,P_2\}+\nu L_3^2.\end{gather}

\newpage
 \begin{center}Table 2.   Inverse masses, potentials  and the related integrals of motion defined up to arbitrary coefficients.\end{center}
\begin{tabular}{c c c c}
\hline
\vspace{1.5mm}№&$f$&$V$&\text{Integrals of motion}\\
\hline\\

1\vspace{1.5mm}&$r^2$&$c\frac{r^2}{x_3^2}$&$\begin{array}{c}\ L_3,
\{L_1,L_2\}+4c\frac{x_1x_2}{x_3^2}\end{array}$\\
2\vspace{1.5mm}&$x_3^2$&$\frac{cx_3^2}{\nu^2 r^2+\nu((\mu+1)x_3^2+\mu x_1^2+x_2^2)+\mu x_3^2}$&$\{P_1,K_1\}+\mu
\{P_2,K_2\}+4\nu\frac{r^2}{x_3^2}V$\\
3\vspace{1.5mm}&$x_3^2$&$\begin{array}{c}c_1\frac{x_3^2}{x_1^2}+c_2\frac{x_3^2}{x_2^2}
+
c_3\frac{bx_1^2+ax_2^2}{x_3^2}\\+c_4\frac{x_3^2}{2(ax_2^2-bx_1^2)+(a-b)x_3^2}\end{array}
$&$
\begin{array}{c}\{P_1,K_1\}+(b+1-a)\{P_2,K_2\}\\+2(a-1)J_3^2+
4c_1\frac{ax_2^2+x_3^2}{x_1^2}
\\+4c_2\frac{(bx_1^2+(b-a+1)x_3^2}{x_2^2}+4c_3\frac{ax_1^2+b(b-a+1)x_2^2}{x_3^2}\\+4c_4
\frac{(2ax_2^2+x_3^2)}{a(x_3^2+2x_2^2)+b(x_3^2+2x_1^2)}\end{array}$\\
4\vspace{1.5mm}&$x_2^2$&$c_1\frac{x_2^2}{x_3^2}+c_2\frac{x_1}{\tilde r}$&$\begin{array}{c}\{P_1,D\}-\{P_3,L_2\}-2c_2\frac{1}{\tilde r}+4c_1\frac{x_1}{x_3^2},\\
\{K_1,D\}-\{K_3,L_2\}-2c_2\frac{r^2}{\tilde r}+4c_1\frac{r^2x_1}{x_3^2}\end{array}$\\
5\vspace{1.5mm}&$x_2^2$&$c\frac{x_2^2}{x_3^2}$&$\begin{array}{c}\{P_1,D\}-\{P_3,L_2\}
+4c\frac{x_1}{x_3^2}, \ K_1,\\
\{K_1,D\}-\{K_3,L_2\}+4c\frac{r^2x_1}{x_3^2}, \ P_1\end{array}$\\
6\vspace{1.5mm}&$x_2^2$&$c\frac{x_1}{\tilde r}$&$\begin{array}{c}\{P_1,D\}-\{P_3,L_2\}-2c\frac{1}{\tilde r}, \ K_3,\\
\{K_1,D\}-\{K_3,L_2\}-2c\frac{r^2}{\tilde r},\ P_3\end{array}$\\
7\vspace{1.5mm}&$x_3^2$&$c_1\frac{x_3^2x_2}{\tilde r x_1^2}+c_2\frac{x_3^2}{x_1^2}$&$\begin{array}{l}\{L_3,P_1\} +4c_2\frac{x_2}{x_1^2}+2c_1\frac{2x_2^2+x_1^2}{\tilde r x_1^2},\\\{L_3,K_1\} +4c_2\frac{x_2r^2}{x_1^2}+2c_1\frac{(2x_2^2+x_1^2)r^2}{\tilde r x_1^2}\end{array}$\\
8\vspace{1.5mm}&$x_3^2$&$c\frac{x_3^2}{x_1^2}$&$\begin{array}{l}\{L_3,P_1\} +4c\frac{x_2}{x_1^2}, \ \{L_3,K_1\} +4c\frac{x_2r^2}{x_1^2},\\ \{P_1,K_1\}+\frac{4cr^2}{x_1^2},\ P_2, \ K_2
\end{array}$\\
9\vspace{1.5mm}&$x_3^2$&$c\frac{\tilde r^2}{r^2}$&$\begin{array}{c}\{K_1,P_2\}+4c\frac{x_1x_2}{r^2}, \ \{K_2,P_1\}+4c\frac{x_1x_2}{r^2},\\  \{P_1,K_1\}+4c\frac{x_1^2}{r^2}, \ L_3,\ \{P_2,K_2\}+4c\frac{x_2^2}{r^2} \end{array}$\\
10\vspace{1.5mm}&$x_3^2$&$c\frac{x_3^2}{r^2}$&$\begin{array}{c}\{P_2,K_1\}
-4c\frac{x_1x_2}{r^2},\ \{P_1,K_2\}-4c\frac{x_1x_2}{r^2}, L_3,\\ \{P_1,K_1\}-\frac{4cx_1^2}{r^2},\  \{P_2,K_2\}-\frac{4cx_2^2}{r^2}\end{array}$\\
11\vspace{1.5mm}&$\tilde r^2$&$c_1e^{-2\varphi}\frac{r^2+x_3^2}{\tilde r^2}+c_2e^{-\varphi}\frac{x_3}{\tilde r}$&$\begin{array}{l}\{P_3,(L_3+ D)\}-4c_1e^{-2\varphi}\frac{x_3}{\tilde r^2}-2c_2e^{-\varphi}\frac1{\tilde r},\\\{K_3,(L_3+ D)\}-4c_1e^{-2\varphi}\frac{r^2 x_3}{\tilde r^2}-2c_2e^{-\varphi}\frac{r^2}{\tilde r}\end{array}$\\
12\vspace{1.5mm}&$\tilde r^2$&$c_1e^{2\varphi}\frac{r^2+x_3^2}{\tilde r^2}+c_2e^\varphi\frac{x_3}{\tilde r}$&$\begin{array}{l}\{P_3,(L_3- D)\}+4c_1e^{2\varphi}\frac{x_3}{\tilde r^2}+2c_2e^{\varphi}\frac1{\tilde r},\\\{K_3,(L_3- D)\}+4c_1e^{2\varphi}\frac{r^2x_3}{\tilde r^2}+2c_2e^{\varphi}\frac{r^2}{\tilde r}\end{array}$\\
13\vspace{1.5mm}&$\tilde r^2$& $c\frac{x_3}r$&$\begin{array}{c}\{P_3,K_3\}+4c\frac{x_3}{ r},\ \{P_3,D\}-\frac{2c_2}{r},\\
\{K_3,D\}-\frac{2c_1 r^2}{\tilde r}-{2c_2}{ r},  \ L_3\end{array}$\\
14\vspace{1.5mm}&$\tilde r^2$&$\frac{\tilde r^2}{x_3^2}$&$\begin{array}{c}P_3^2+\frac1{x_3^2},\ K_3^2+\frac{r^4}{x_3^2}, \ L_3 \end{array}$\\
15\vspace{1.5mm}&$\tilde r^2$&$c$&$P_3, \ L_3,\ D, \ K_3$\\
16\vspace{1.5mm}&$x_3^2$&$c$&$P_1,\  P_2, \ K_1,  \ K_2, D, L_3$\\
17\vspace{1.5mm}&$r^2$&$c$&$L_1,\  L_2, \ L_3,   D$\\
\hline\hline
\end{tabular}

\vspace{1.5mm}

The corresponding matrix $M$  is degenerated, and its nonzero  entries  (\ref{deq7})
take the following form:
\begin{gather*}M^{11}=2(x_3^2+(1+\nu)x_2^2),\ M^{12}=-2(\mu+\nu)x_1x_2,\ M^{21}=-2(1+\nu)x_1x_2,\\ M^{22}=2(\mu+\nu)x_1^2+\mu x_3^2, \ M^{13}=-2x_1x_3, \
M^{23}=-2\mu x_2x_3.\end{gather*}

The related equations (\ref{deq5}) are solved by $f=x_3^2$, and the corresponding equations  (\ref{mmmu3}): take the following form:
\begin{gather}\la{new1}\begin{split}&(ax_2^2+x_3^2)V_1-bx_1x_2V_2=x_3^2\eta_1,\\&
(bx_1^2+(b+1-a)x_3^2)V_2-bx_1x_2V_1=x_3^2\eta_2\end{split}\end{gather}
where $a=1+\nu$ and $b=\mu+\nu$. Notice that the third component of equations  (\ref{mmmu3}) in our case is a consequence of the system (\ref{new1}) since $\eta$ should satisfy the condition $x_aK_a=0$.

By definition potential $V$ should be scale invariant and so can be treated as a function of two scale invariant variables $y_1=\frac{x_1}{x_3}$ and $y_2=\frac{x_2}{x_3}$. The  system  (\ref{new1}) is compatible provided the following second order equation for $V$ is satisfied:
\begin{gather}\la{new2}(ay_2^2-by_1^2+a-b)V_{ y_1y_2}+y_1y_2(aV_{y_1y_1}-bV_{y_2 y_2})+3(ay_2V_{y_1}-by_1V_{y_2}) = 0.\end{gather}

The generic solution of the system (\ref{new1}) can be found in a a closed  form only for special combinations of parameters $a$ and $b$, namely, $a=b$ and $a=0$ (or $b=0$ which is the same). They are represented in Items 8-10 of Table 1. For $a$ and $b$ arbitrary we were able to find only particular solutions which are presented in Items 2 and 3 of Table 2. The arbitrary solutions are denoted as $\tilde V$ and $\tilde \eta$ and represented formally in Item 11 of Table 1.

  Analogously it is possible to solve the equations corresponding to the remaining inequi-valent matrices $M$. Doing this it is necessary  to extend matrices $N$ by adding   special matrices  $\tilde N$ (refer to equation (\ref{deq111})) which keeps $M$ degenerated. The obtained in this way results are collected in two tables one of which includes systems defined up to arbitrary functions, the other one includes arbitrary coefficients. The presen-ted list of PDM systems admitting second order integrals of motion is complete up to rotation transformations.

In the tables  $F(.), G(.)$ and  $R(.)$ are arbitrary functions of the arguments specified in brackets, $c, c_1, c_2, \mu$ and $\nu$ are arbitrary real parameters  $\varphi$ and $\theta$ are Euler angles, $ r^2=x_1^2+x_2^2+x_3^2,\ \tilde r^2=x_1^2+x_2^2, $ $  P_a, K_a, D$ and $L_3$ are operators collected in (\ref{QQ}), and the summation is imposed over the repeating indices  $a$ by values 1, 2 and 3. The symbol $\{A,B\}$ denotes the anticommutator of operators  $A$ and $B$, i.e.,
$\{A,B\}=AB+BA.$ For the systems fixed in the three latest items of Table 2 the second order integrals of motion are reduced to bilinear combinations of the first order symmetries indicated there.

 \section{Concluding remarks}

We find all inequivalent PDM Schr\"odinger equations  which are scale invariant and admit second order integrals of motion. In the classification  tables thirteen  versions of such equations are collected, part of which include arbitrary parameters which can be treated as coupling constants (see Table 2). The remaining equations presented in Table 1 are defined up to arbitrary functions of the reduced numbers of independent variables.

Notice that the systems presented in Items 7, 12 of Table 1 and Items 1, 3, 7, 8, 12, 13 are invariant w.r.t. the rotations around the third coordinate axis.

Since all found  hamiltonians by definition commute with the dilatation generator $D$, the related Schr\"odinger equations are integrable if they admit two second order integrals of motion, and superintegrable if the number of such integrals of motion is more extended. In accordance with the tables there exist a rather extended number of integrable and superintegrable PDM systems in the considered class.

Thus we present a number of new integrable PDM Schr\"odinger equations. The main value of this result is its completeness, since we find all inequivalent equations in the considered class. It can be considered as the first step in the description of all PDM quantum systems which admit second order integrals of motion and are invariant w.r.t. at least one parametrical Lie group.

A natural next step is construction of exact solutions of the obtained integrable and superintagrable systems and the description of analogous systems invariant w.r.t. all inequivalent one parametric lie groups. The latter task is not too cumbersome since the number of such groups is not too extended and is equal to four \cite{NZ}.  Let us remind that the exact solutions for the PDM Schr\"odinger equations admitting at least five parametric symmetry groups were found in
 \cite{AN!}.


\begin{thebibliography}{99}

 \bibitem{FN1}V. I. Fushchich, and A. G. Nikitin. Symmetries of equations of quantum mechanics. New York: Allerton Press (1994).

\bibitem{FN11}W. I Fushchych and A. G. Nikitin, On the new invariance groups of the Dirac and Kemmer-Duffin-Petiau equations, Lettere al Nuovo Cimento 19, 347-352 (1977).

\bibitem{N22}  A. G. Nikitin, The maximal ”kinematical” invariance group for an arbitrary potential
revised, Journal of Mathematical Physics, Analysis, Geometry 14, 519-531 (2018).

\bibitem{N32}  A. G. Nikitin, Symmetries of Schr\"odinger equation with scalar and vector potentials,
J. Phys. A: 53, 455202 (2020).

\bibitem{N42}  A. G. Nikitin, Symmetries of the Schr\"odinger-Pauli equation for neutral particles, J.
Math. Phys. 62, 083509 (2021).

\bibitem{N52}  A. G, Nikitin Symmetries of the Schr\"odinger-Pauli equations for charged particles
and quasirelativistic Schr\"odinger equations,  J. Phys. A 55, 115202 (2022).

  \bibitem{Nied}
U. Niederer, The maximal kinematical invariance group of the free
Schr\"odinger equations,
 Helv. Phys. Acta, {45}, 802--810 (1972).


\bibitem{And}
R. L. Anderson, S. Kumei, C. E. Wulfman,  Invariants of the
equations of wave mechanics. I., Rev. Mex. Fis. { 21},  1--33
(1972).


\bibitem{Boy}
C. P. Boyer, The maximal kinematical invariance group for an
arbitrary potential, Helv. Phys. Acta { 47}, 450--605 (1974).

\bibitem{wint1}P.  Winternitz, J. Smorodinsky, M. Uhli\u{r} and I. Fri\u{s},
 Symmetry groups in classical and quantum mechanics,
 { Sov. J. Nucl. Phys.} { 4}, 444--450 (1967).


\bibitem{BM}A. Makarov, J. Smorodinsky, Kh. Valiev and P. Winternitz
 A systematic search for non-relativistic systems with
dynamical symmetries, { Nuovo Cim. A} { 52},  1061--1084 (1967)


\bibitem{Mar}Ian Marquette and Pavel Winternitz. Higher order quantum superintegrability: a new Painleve
conjecture. Integrability, Supersymmetry and Coherent States. Springer, Cham, pp. 103-131.
 (2019).
\bibitem{AGN1} A. G. Nikitin, Higher-order symmetry operators for Schr\"odinger equation. In CRM Proceedings
and Lecture Notes (AMS), 37 , pp. 137–144 (2004).

\bibitem{Roz}Oldwig von Roos, Position-dependent effective masses in semiconductor theory,
Phys. Rev. B { 27}, 7547 (1983).
\bibitem{7} A. de Saavedra, F. Boronat, A. Polls and A. Fabrocini, Effective mass of one He 4 atom in liquid He 3, Phys. Rev. B { 50}, 4248 (1994).

\bibitem{3} P. Harrison, { Quantum Wells, Wires and Dots} (Wiley,
New York, 2000).

\bibitem{Miller1}
R. Heinonen, E. G. Kalnins, W. Miller Jr and E. Subag,  Structure relations and Darboux contractions for 2D 2nd order superintegrable systems, SIGMA { 11}, 043 (2015).

\bibitem{Miller2}  B. K.   Berntson, E. G. Kalnins, and W. Miller Jr. Toward classification of 2nd order superintegrable systems in 3-dimensional conformally flat spaces with functionally linearly dependent symmetry operators, SIGMA: Symmetry, Integrability and Geometry: Methods and Applications, 16, 135. (2020).

\bibitem{Bala2} A. Ballesteros, A. Enciso, F. J. Herranz, O. Ragnisco and D.
Riglioni, Superintegrable oscillator and Kepler systems on
spaces of nonconstant curvature via the St\"ackel transform, {
SIGMA} { 7} 048 (2011).

\bibitem{Rag1}O. Ragnisco and D. Riglioni,   A Family of Exactly Solvable Radial Quantum Systems on Space of Non-Constant
Curvature with Accidental Degeneracy in the Spectrum, { SIGMA}
{ 6}  097 ( 2010).

\bibitem{N62} A. G. Nikitin, Superintegrable and shape invariant systems with position dependent mass,
     J. Phys. A: Math. Theor. 48 335201 (2015).

 \bibitem{NZ}  A. G. Nikitin and  T. M.  Zasadko, Superintegrable systems with
position dependent mass, { J.
Math. Phys.} { 56} 042101 (2015).

\bibitem{AGN}A. G. Nikitin, Kinematical invariance groups of the 3d Schrödinger equations with position dependent masses,
J.  Math. Phys. 58, 083508 (2017).

\bibitem{AGN0} A. G. Nikitin, Group classification of systems of non-linear reaction-diffusion equations with general diffusion matrix. I. Generalized Ginzburg-Landau equations, J. Math. Analysis and Appl. 324, 615-628 (2007).


\bibitem{NNN} A. G. Nikitin, Group classification of systems of nonlinear reaction-diffusion equations with triangular diffusion matrix,
Ukr. Math. J. 59, 439-458    (2007).

\bibitem{NNNN}A. G. Nikitin and V. I. Fushchich, Equations of motion for particles of arbitrary spin invariant under the Galileo group,
Theor. and Math. Phys. 44, 584-592 (1980)


\bibitem{Rom}O. O. Vaneeva, R. O. Popovych and C. Sophocleous, Equivalence transformations in the study of integrability, Physica Scripta, 89, 038003 (2014).

     \bibitem{Kil}A. G.  Nikitin, Generalized Killing tensors of arbitrary
     valence and order, {  Ukr. Math. J.}
     { 43} 734--743 (1991).

\bibitem{AN!}A.G. Nikitin, Exact solvability of PDM systems with extended Lie symmetries,
Proceedings of the Institute Mathematics of the Nat. Acad. Sci. of Ukraine 16, No 1, 1--18 (2019).

\end{thebibliography}
\end{document}